\documentclass{aa}  

\usepackage{graphicx}
\usepackage{txfonts}
\usepackage{natbib}
\usepackage{rotating}

\bibpunct{(}{)}{;}{a}{}{,}

\newcommand{\kms}{km\,s$^{-1}$}

\newcommand{\vsini}{$v\sin{i}$}
\newcommand{\betuma}{$\beta$~UMa}
\newcommand{\tetleo}{$\theta$~Leo}

\begin{document}

   \title{Detection of ultra-weak magnetic fields in Am stars: \\$\beta$ UMa and $\theta$ Leo}


   \author{A. Blaz\`ere\inst{1,2,3}
\and P. Petit\inst{2,3}
\and F. Ligni\`eres\inst{2,3}
\and M. Auri\`ere\inst{2,3}
\and J. Ballot\inst{2,3}
\and T. B\"ohm\inst{2,3}
\and C.P. Folsom\inst{2,3,4}
\and M. Gaurat\inst{2,3}
\and 
\\L. Jouve\inst{2,3}
\and A. Lopez Ariste\inst{2,3}
\and C. Neiner\inst{1}
\and G.A .Wade\inst{5}
          }

\institute{LESIA, Observatoire de Paris, PSL Research University, CNRS, Sorbonne Universit\'es, UPMC Univ. Paris 06, Univ. Paris Diderot, Sorbonne Paris Cit\'e, 5 place Jules Janssen, 92195 Meudon, France\\
              \email{aurore.blazere@obspm.fr}
\and
              Universit\'e de Toulouse, UPS-OMP, Institut de Recherche en Astrophysique et Plan\'etologie, Toulouse, France  
              \and
   CNRS, Institut de Recherche en Astrophysique et Plan\'etologie, 14 Avenue Edouard Belin, F-31400 Toulouse, France
                \and
                IPAG, UJF-Grenoble 1/CNRS-INSU, UMR 5274, 38041 Grenoble, France
                \and
Department of Physics, Royal Military College of Canada, PO Box 17000 Station Forces, Kingston, ON K7K 0C6, Canada
             }


 
  \abstract
   {An extremely weak circularly polarized signature was recently discovered in spectral lines of the chemically peculiar Am star Sirius A. A weak surface magnetic field was proposed to account for the observed polarized signal, but the shape of the phase-averaged signature, dominated by a prominent positive lobe, is not expected in the standard theory of the Zeeman effect.}
   {We aim at verifying the presence of weak circularly polarized signatures in two other bright Am stars, $\beta$ UMa and $\theta$ Leo, and investigating the physical origin of Sirius-like polarized signals
further.}
    {We present here a set of deep spectropolarimetric observations of $\beta$ UMa and $\theta$ Leo, observed with the NARVAL spectropolarimeter. We analyzed all spectra with the Least Squares Deconvolution multiline procedure. To improve the signal-to-noise ratio and detect extremely weak signatures in Stokes V profiles, we co-added all available spectra of each star (around 150 observations each time). Finally, we ran several tests to evaluate whether the detected signatures are consistent with the behavior expected from the Zeeman effect.}
    {The line profiles of the two stars display circularly polarized signatures similar in shape and amplitude to the observations previously gathered for Sirius A. Our series of tests brings further evidence of a magnetic origin of the recorded signal.}
    {These new detections suggest that very weak magnetic fields may well be present in the photospheres of a significant fraction of intermediate-mass stars. The strongly asymmetric Zeeman signatures measured so far in Am stars (featuring a dominant single-sign lobe) are not expected in the standard theory of the Zeeman effect and may be linked to sharp vertical gradients in photospheric velocities and magnetic field strengths.}
   

   \maketitle
%

\section{Introduction}

About 5\% to 10\% of hot stars (stars with O, B, and A spectral types) are found to be strongly magnetic with a longitudinal magnetic field strength in excess of 100 G (e.g., \citealt{wade13,auriere07}), which is generally associated with a simple and stable field geometry (e.g., \citealt{lueftinger10,silvester14}). However, the physical origin and even some basic properties of these magnetic fields are still poorly understood. The current paradigm, the fossil field hypothesis, describes this magnetism as the remnant of magnetic field accumulated or produced during an early phase of stellar life. In this conceptual framework, magnetic fields observed in these stars today are proposed to result from a seed field in the molecular cloud from which the star was formed, rather than being currently produced by an active dynamo as in the Sun. This initial field may also have been amplified during the early phases of the evolution of the star, when it was temporarily surrounded by an extended convective envelope hosting a global dynamo. In practice, the fossil field theory leaves many basic questions unanswered, such as the precise origin of this magnetism and its low incidence among intermediate-mass and massive stars. However, it is strongly supported by many of their observational properties (e.g., \citealt{braithwaite15}).

Recently, a longitudinal magnetic field much weaker than any previous detection in intermediate-mass stars has been discovered in the early A star Vega \citep{lignieres09}. The spectropolarimetric time series was interpreted in terms of a surface magnetic field distribution using the Zeeman-Doppler Imaging technique (ZDI, \citealt{petit10}), unveiling a peak local field strength of about 7~G \citep{petit14b}. The results of that study support the view that Vega is a rapidly rotating star seen nearly pole-on, and the reconstruction of the magnetic topology at two epochs revealed a magnetic region of radial field orientation, closely concentrated around the rotation pole. Vega may well be the first confirmed member of a much larger, as yet unexplored, class of weakly magnetic hot stars. Weak magnetic fields of the same kind were also searched for in two normal B stars, $\gamma$ Peg \citep{neiner14b} and $\iota$ Her \citep{wade14}, although no magnetic fields were detected in both stars with a precision of 0.3-0.4~G. However, \cite{wade14} demonstrate that, if a large-scale magnetic field identical to the ZDI magnetic geometry of Vega existed in $\gamma$ Peg and $\iota$ Her, no detection would be expected at this level of accuracy given the spectral line properties of both B-type targets. 

The only other example of a weak Stokes V detection in spectral lines of an intermediate-mass star has been reported for the bright Am star Sirius A \citep{petit11}. For this object, however, the polarized signature observed in circular polarization is not a null integral over the width of the line profile, as expected in the usual descriptions of the Zeeman effect. Instead, the Stokes V line profile exhibits a positive lobe dominating the negative one (in amplitude and integrated flux). The interpretation of a Zeeman origin was favored by Petit et al., in particular after excluding the possibility of an instrumental crosstalk from linear to circular polarization. However, the abnormal shape of the polarized profile remained a puzzle and required further investigation. 

The motivation to progress on this topic is strong because the discovery of a new, potentially widespread class of weakly magnetic A stars offers important new information about the dichotomy between strong and weak magnetic fields in tepid stars. In an attempt to interpret this division, \cite{auriere07} proposed a scenario based on the stability of a large scale magnetic configuration in a differentially rotating star, leading to estimating a critical field strength above which magnetic fields can remain stable on long time scales, while magnetic fields below this limit would likely be destroyed by the internal shear. More detailed models including 2D and 3D numerical simulations \citep{jouve15,gaurat15} tend to confirm the existence of a critical field in such configurations, where the pre-main sequence contraction is a possible way to force differential rotation.  On the other hand, the magnetic dichotomy might simply be the result of two different magnetic field generation processes. \cite{braithwaite13} propose that Vega-like magnetic stars are the result of the slow evolution of magnetic configurations characterized by weak initial magnetic helicity and argue that it should be widespread among most intermediate-mass and massive stars. Meanwhile, \cite{ferrario09} propose that the small fraction of strong magnetic fields could be produced in early stellar merging events.

In the rest of this paper, we first present the two bright Am stars selected for this study. We then present the observations and the analysis methods used. A series of tests was performed to constrain the physical origin of the recorded polarimetric signatures further, and finally we discuss our results in the broader context of weakly magnetic star of intermediate mass.

\section{Selected targets}

\begin{table}[h]
\label{parameter}
\caption{Fundamental parameters of $\beta$ UMa and $\theta$ Leo}
\centering
\begin{tabular}{c c c}
\hline
  & $\beta$ UMa & $\theta$ Leo\\
\hline
\hline
   Spectral type & A1V & A2V \\
   $T_{eff}$    & 9480K$\pm$10K$^{a}$ & 9280$\pm$10K$^{a}$  \\
   log g   &   3.82$^b$   & 3.65$^c$ \\
   Mass & 2.64$\pm$0.01$M_{\odot}^{a}$ & 2.94$\pm$0.2$M_{\odot}^{a}$\\
   Radius & 3.021$\pm$0.038 $R_{\odot}^{d}$ & 4.03$\pm$ 0.10 $R_{\odot}^{e}$\\
   \vsini & 46$\pm$3 km/s$^{f}$ & 23$\pm$3 km/s$^{f}$\\
   L$_{\odot}$ & 72$\pm$11$^{a}$ & 127$\pm$13$^{a}$\\
   Frac. age & 0.778$^{a}$ & 0.943$^{a}$ \\
   Metallicity & -0.03$^{g}$ & -0.13$^{g}$\\
\hline

\multicolumn{2}{l}{$^{a}$ \cite{zorec12}}&$^{b}$\cite{allendeprieto99}\\
\multicolumn{2}{l}{$^{c}$ \cite{adelman15b} }& $^{d}$ \cite{boyajian12}\\
\multicolumn{2}{l}{$^{e}$ \cite{maestro13}} & $^{f}$ \cite{royer02}\\
\multicolumn{2}{l}{$^g$ \cite{anderson12}} & \\

\end{tabular}
\end{table}

Here, we present the results of deep spectropolarimetric campaigns carried out for two bright Am stars in which magnetic fields were previously undetected \citep{auriere10}. Am stars are chemically peculiar stars exhibiting overabundances of iron-group elements such as zinc, strontium, zirconium, and barium and deficiencies of a few elements, particularly calcium and scandium. Most Am stars also feature low projected rotational velocities, as compared to normal A stars \citep{abt09}. The targets of this study are $\beta$ Ursa Majoris (HD 95418) and $\theta$ Leonis (HD 97633). Abundances measured for $\beta$ UMa place this star among targets featuring weak Am characterictics with noticeable overabundance in V{\sc ii}, Mn{\sc ii}, Ni{\sc i}, Ni{\sc ii}, Zn{\sc i}, Sr{\sc ii}, Y{\sc ii}, Zr{\sc ii}, and Ba{\sc ii} and underabundances in He{\sc i}, C{\sc i}, C{\sc ii}, and Sc{\sc ii} (for more details see \citealt{adelman11}). The source $\theta$ Leo is also on the weak side of Am abnormality, with large reported overabundance in S{\sc ii}, V{\sc ii}, Cr{\sc ii}, Sr{\sc ii}, Y{\sc ii}, and Zr{\sc ii} and Ba{\sc ii} and underabundance in A{\sc ii}, Ca{\sc ii}, Sc{\sc ii}, Mn{\sc ii}, and Ni{\sc i}, \citep{adelman15b}. 

The fundamental parameters of both targets are presented in Table~\ref{parameter}. The two objects are early A-type targets. Both of them benefit from an interferometric estimate of their radius, which is distinctly larger than the radius of main sequence stars of similar spectral types. Accordingly, their surface gravities are found to be below main sequence values. High luminosity values complete this picture, confirming that both targets are already on their way off the main sequence. Using evolutionary models matching the position of both stars in the H-R diagram, \citet{zorec12} find that the fractional age on the main sequence of \betuma\ and \tetleo\ are equal to 0.778  and 0.943, respectively, giving further support to the idea that both stars have completed most of their path on the main sequence. That \betuma\ is reported to belong to the Ursa Majoris association gives another constraint on the age, which is estimated to be around 500~Myr for this group of stars \citep{monier05}. Based on Spitzer measurements of IR excess, \cite{ballering13} attribute ages of 310~Myr and 500~Myr to \betuma\ and \tetleo, respectively, which is too young to be reconciled with other stellar parameters, but may provide an additional hint that \tetleo\ is more evolved than \betuma. 

The projected rotational velocities estimated for both stars are fairly typical of values reported for Am stars \citep{abt09}. In the absence of any direct estimate of the rotation period of our targets, it cannot be determined whether the higher \vsini\ value reported for \betuma\ is linked to a faster rotation or higher inclination angle.

\section{Data analysis}

Data were taken with the NARVAL spectropolarimeter (\citealt{auriere03}, \citealt{silvester12}) in operation at the two-meter Bernard Lyot
Telescope (TBL) at the summit of Pic du Midi Observatory in the French
Pyr\'en\'ees. This
high resolution spectropolarimeter is specially designed and
optimized to detect stellar magnetic
fields through the polarization they generate in photospheric spectral
lines. The polarimetric unit is mounted at the Cassegrain focus of the
telescope and allows two orthogonal states of a given polarization
(circular or linear) to be recorded throughout the entire optical
domain, thanks to the high achromaticity of its polarimetric optics.
The upper part of the polarimeter contains the guiding camera, an atmospheric dispersion corrector, and
a calibration wheel.  Following that, the main polarimetric device constitutes three Fresnel rhomb retarders (two half-wave rhombs that can rotate about the optical axis and one fixed quarter-wave rhomb), which are used to perform the polarimetric analysis. The light emerging from the retarders is sent
to a Wollaston prism, consisting of two orthogonal calcite prisms that
are cemented together, acting as a polarizing beamsplitter.

The two beams of light emerging from the beamsplitter are transmitted
by some 30 m of optical fiber to the bench-mounted spectrograph, where an image slicer converts the circular image of the fiber head into a pseudo-slit shape.. The spectrograph
provides complete coverage of the optical spectrum from 3700 to 10500
$\AA$ on 40 echelle orders with a spectral resolution of about 65000
in polarimetric mode. The spectrograph unit contains a double set
of high-reflectance collimators cut from a single 680 mm parabolic
mirror with a focal length of 1500 mm. The grating is a 79 gr/mm
monolithic grating with dimensions of 200 by 400 mm, and the
cross-dispersion is achieved by a high dispersion prism. The camera
lens is a fully dioptric f/2 388 mm focal length lens with a 210 mm
free diameter. The spectrograph thermal stability is kept to within
0.1~K, thanks to the use of a double-layer thermal enclosure.
\\

All data used in the present paper are collected in the polarimetric
mode measuring Stokes V (circular polarization).To minimize systematic
errors, one complete Stokes V sequence consists of four successive
subexposures taken with the half-wave rhombs oriented at different
azimuths. This follows the method of \cite{semel93} to reduce the
amplitude of possible spurious signatures of instrumental origin. This strategy also provides a
strong test to discard the possibility of a spurious signal by
computing a ``null'' spectrum. This is calculated from a different
combination of the four subexposures constituting the polarimetric
sequence \citep{donati97}, and it should not display any signal. This ``null'' check parameter is automatically produced for each Stokes V sequence.The
data are reduced by Libre-Esprit, a dedicated and fully automated
software \citep{donati97} specifically developed for reducing echelle
spectropolarimetric data and optimized for NARVAL. Libre-Esprit proceeds in three steps: the first
stage consists of performing a geometrical analysis from a sequence of
calibration exposures; the position and shape of orders is derived
from a mean flat field image, while the details of the wavelength to
pixel relation along and across each spectral order is obtained from comparison frames obtained from a ThAr lamp and a Fabry-Perot setup. The second step performs
spectrum optimal extraction \citep{horne86,marsh89}, using
the geometrical information derived in step 1. A last step consists of
refining the wavelength calibration using telluric lines recorded in
the reduced spectrum, therefore reaching a radial velocity accuracy
close to 30 $m.s^{-1}$ \citep{moutou07}. Spectra processed with
Libre-Esprit include the flux and polarization information, as well as
the ``null'' spectrum computed from two different combinations (dubbed ``Null1'' and ``Null2'' in our plots) and error bars at each wavelength point in the
spectrum.

The source $\beta$ UMa was observed in March/April 2010 and March/April 2011 for
a total of 149 spectra. For its part, $\theta$ Leo was observed in
January/March/April 2012, March/April 2013, and May/June 2014 for a
total of 171 spectra (see Table~\ref{obs} for the detailed distribution of observations among individual nights). For each star, the exposure time was adjusted to reach a peak S/N throughout the Stokes V spectrum between 1,000 and 2,000 per 1.8 $km.s^{-1}$ bin, depending on weather conditions. These relatively high values are safely away from the saturation level of the EEV detector used in fast readout mode.

In the absence of any detectable polarized signatures in individual
spectral lines of $\beta$ UMa and $\theta$ Leo, we apply the
well-known and commonly used Least Squares Deconvolution (LSD)
procedure (\citealt{donati97}, \citealt{kochukhov10b}) to each
spectrum of both stars. This method is a cross-correlation technique
for computing average pseudo-line profiles from a list of spectral
lines in order to get a multiplex increase in the S/N. This powerful technique, based upon several rough approximations (additive line profiles,
wavelength-independent limb-darkening, self-similar local profile
shape, weak magnetic fields), makes use of the possibility of describing stellar spectra as a
line pattern convolved with an average pseudo-line profile. Here, we
choose to compute the LSD Stokes I and V pseudo-profiles for all available
photospheric lines. Our line lists are taken from the VALD atomic data
base \citep{piskunov95, kupka99} using the respective effective
temperature and log g of both stars (Table~\ref{parameter}). Our line
lists are extracted using $T_{eff}=9,500K$ and log g=4.0 for $\beta$
UMa and $T_{eff}=9,250K$ and log g=3.5 for $\theta$ Leo. We reject the
lines whose depth is less than 1\% of the continuum. By doing so we obtain a mask of 1,173 lines and 1,133 lines for $\beta$
UMa and $\theta$ Leo, respectively. Then, we adjust the depth of the lines in the
mask to fit the observed line depths. To reduce the noise per spectral bin
further and then reduce the spectral
resolution of LSD line profiles. Instead of the default spectral bin
spanning 1.8 \kms\ at a spectral resolution of 65,000, we used 9 \kms\
for $\beta$ UMa and 5.4 \kms\ for $\theta$ Leo, which leaves us with
about ten velocity bins in the pseudo line profile. With this modification of
the spectral resolution, the additional gain in the S/N is a factor of 2.1 for
$\beta$ UMa and 1.7 for $\theta$ Leo. The nightly averaged S/Ns of the
resulting Stokes V LSD profiles (i.e., the average of the S/Ns of individual profiles) are between 45,000 and 77,000
for $\beta$ UMa and between 25,000 and 48,000 for $\theta$ Leo
(Table~\ref{obs}). The dispersion of the S/N between individual Stokes V sequence of a given night is often the lowest during
nights featuring the highest average S/N, because of the excellent (and stable) sky transparency.

Polarized signals remain undetected in individual LSD Stokes V
pseudo-profiles of our two targets. However, their typical S/N remains far
too low to detect polarized signatures as weak as the one previously
reported for Sirius A \citep{petit11}. To further improve the
S/N, we coadd all available LSD profiles for each
star, resulting in one ``grand average'' pseudo-line profile. This method
was successfully used for Vega (\citealt{lignieres09}, \citealt{petit10}) and Sirius \citep{petit11} to detect signatures with amplitudes as low as about
$10^{-5}$ of the continuum level. To coadd the LSD profiles, we weight
each individual LSD profile proportionally to its squared S/N:

\[w_{i}=S/R_{i}^2/\sum \limits_{\underset{}{i=0}}^n S/R_{i}^2\]

\noindent where $w_{i}$ and $S/R_{i}$ are the weight and S/N of the $i^{th}$ pseudo profile. 

\begin{figure}[H]
\centering
\includegraphics[scale=0.34]{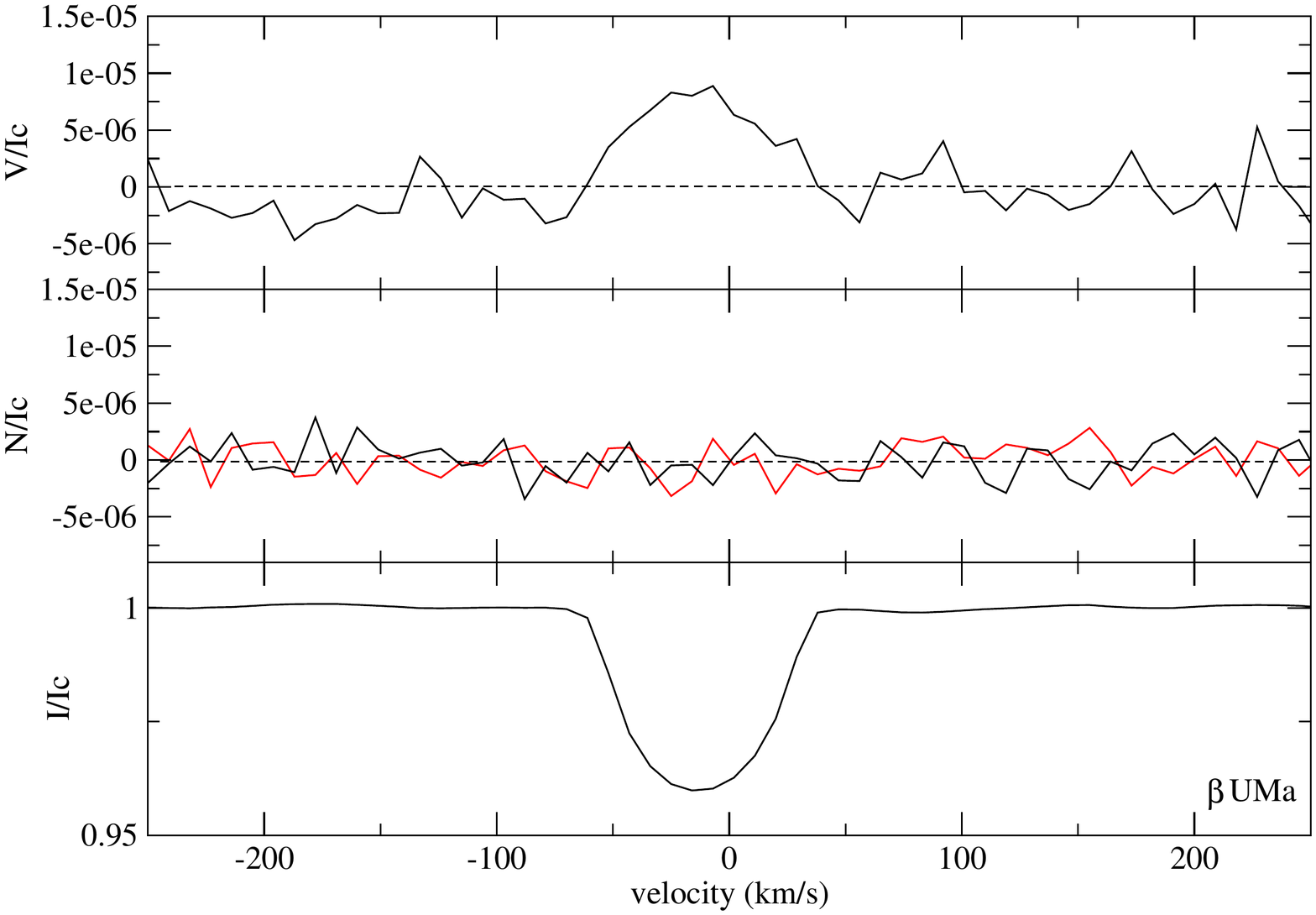}
\includegraphics[scale=0.34]{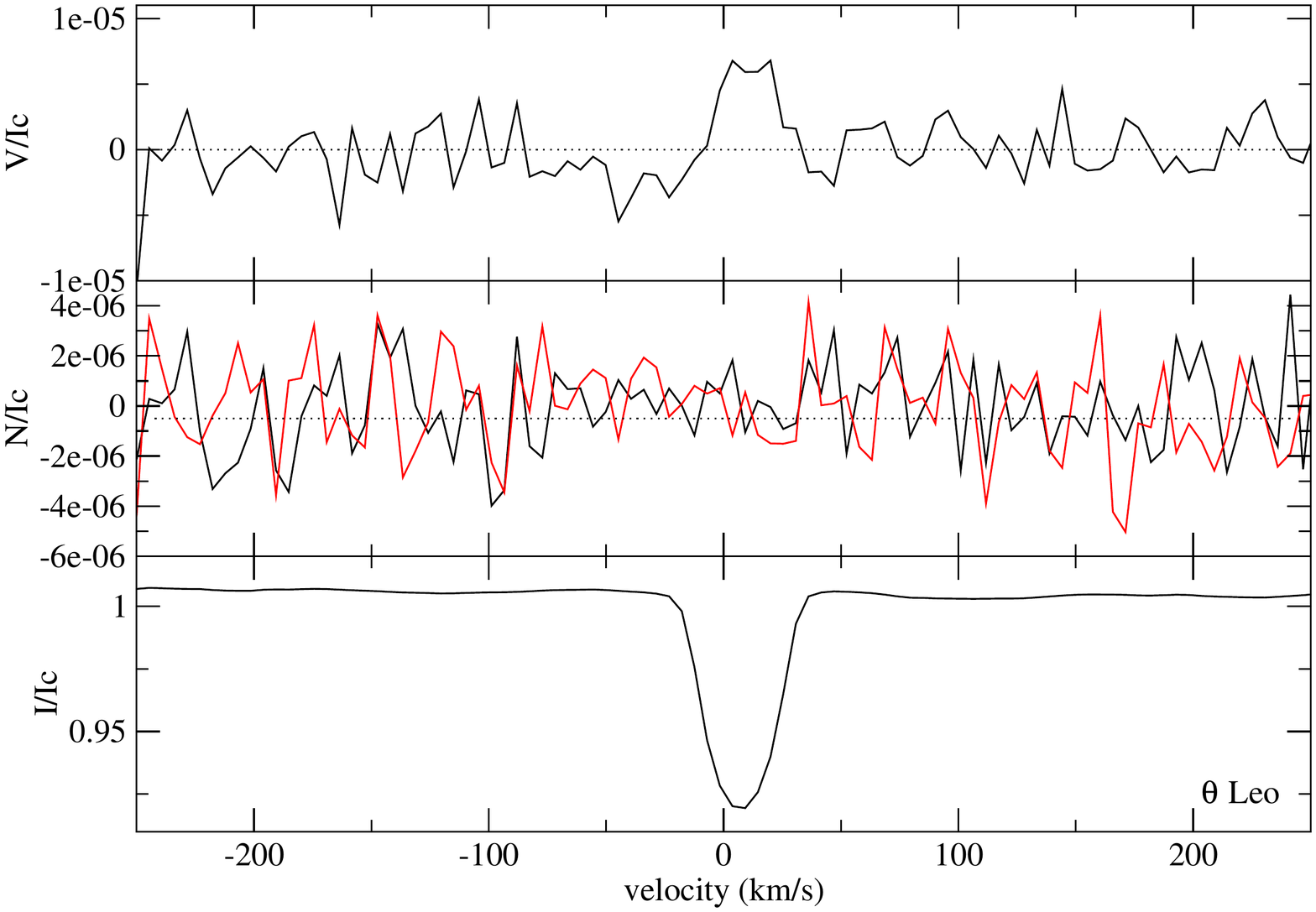}
\caption{Coadded LSD profiles in Stokes I (bottom) and V (top). The two available ``null'' control parameters Null1 and Null2 are shown in the middle panel. Top: $\beta$ UMa observations. Bottom: Same figure for $\theta$ Leo. All profiles are normalized to the continuum level.}
\label{lsd}
\end{figure}

We
choose here to keep all profiles in this process, even those with the
lowest S/N (LSD profiles with S/N lower than 10,000 represent 16
observations for $\beta$ UMa and 2 for $\theta$ Leo), because this
systematic rejection was found to provide us with nothing more than a marginal modification of the
result (and no noticeable improvement). The grand average LSD profiles are presented in
Fig.~\ref{lsd}. With the large number of spectra collected here, the
coaddition of all profiles increases the S/N by a factor $\approx$ 10,
compared to individual profiles. The resulting S/N of the grand average
V profiles is 653,640 for $\beta$ UMa and 512,370 for $\theta$ Leo (using the normalization parameters listed in Table \ref{lsd_parameter}). 

One
limitation of this rough co-addition method is that we average together
observations taken at different rotational phases. In the absence of any known rotation period, we assume our data are distributed over all rotation phases with the same probability. We therefore lose any phase-resolved
information, and the axisymmetric surface structures (i.e., structures symmetric about the spin axis) are the most likely to survive the coaddition process and actually 
contribute to the grand average. This strategy is, however, successful
at reducing the noise level enough to permit the detection of
circularly polarized signatures in both stars, while the null profiles
remain free of any feature above noise level.

Reduced spectra are provided by Libre-Esprit with a normalized
continuum, although the actual resulting continuum typically deviates by up to
15\% from unity, especially in the bluest orders of the spectra. To
test the impact of this imperfect automated processing on the result of our LSD analysis, we
normalized each of the 40 echelle orders for each spectrum (see
\citealt{neiner14}) with the continuum task of IRAF\footnote{Image
Reduction and Analysis Facility, http://iraf.noao.edu/}. The new
normalization improves the S/N of the individual LSD profiles by about 5\%. We
notice that the upgraded normalization changes the resulting
LSD profiles slightly, however the improvement is very marginal, even at the
extremely high S/N of our grand average profiles. In spite of the limited quality of the default continuum normalization, the robustness of
LSD is mainly due to the
large number of lines taken into account in the LSD process, compared
to hotter stars for which the improved normalization is more useful. As a consequence, we simply consider here the spectra normalized 
with Libre-Esprit for consistency with the previous
studies on Vega (\citealt{lignieres09}, \citealt{petit10}) and Sirius
\citep{petit11} in which the default renormalization was used.

\section{Results}

\subsection{LSD profiles with complete line mask}
\label{sect:complete}

The Stokes I, V, Null1 and Null2 co-added LSD profiles of $\beta$ UMa and $\theta$ Leo are shown in Fig.~\ref{lsd}. They display clear Stokes V signatures at the radial velocity of the Stokes I line profiles. The circularly polarized signal observed for both stars covers most of the width of the line and is mostly symmetric about the line centroid. In both cases, a positive lobe dominates the signal. No detectable signal is seen in the Null1 and Null2 control profiles. We computed the detection probability of the Stokes V signal by using the $\chi^2$ test proposed by \cite{donati92}, getting a detection probability of $\sim$100~\% for both stars with a false alarm probability below $10^{-11}$  for $\beta$ UMa and equal to 6.5$\times 10^{-6}$ for $\theta$ Leo. Outside of the stellar lines, we obtained a marginal signal detection for $\beta$ UMa, due to the negative bump in the Stokes V continuum showing up at a radial velocity of around -200 \kms. This continuum feature, not observed for \tetleo, may be due to residuals of line blends \citep{kochukhov10b}. 

We note that the Stokes V signatures detected in the co-added LSD profiles probably stem from a significant fraction of the individual LSD profiles, as various subsets from our complete data set (e.g., observations taken during a given year, see Fig. \ref{year}) display the same signal when co-added separately, although with a higher noise level. The single-epoch subsets are obtained over a timespan that is much longer than the longest possible rotation period of the two targets, so that the co-addition process of many individual rotational phases should result, in all cases, in a filtering of any signatures resulting from nonaxisymmetric magnetic structures.

\subsection{Possible instrumental artifacts at high S/N}

The very high S/N achieved to detect weak polarimetric signatures in intermediate-mass stars raises the question of possible instrumental effects that could contribute to generate spurious signatures in NARVAL Stokes V sequences.  All spectra obtained for our study display a peak S/N below 2,000, and the majority of them are kept below 1,500. At such S/N values, we safely stay away from the saturation regime of the detector (S/N above 2,000 for standard early-type stars). We note that subsets extracted from our complete time-series display consistent signatures, regardless of the S/N of the subset, as highlighted by, e.g., Fig. \ref{year}. In any case, most spurious signatures generated by nonlinear behavior of the detector are expected to show up in the Null1 and Null2 check profiles (especially if the S/N is fluctuating from one subexposure to the next), which is not seen here. 

From an empirical point of view, we stress that the signatures recorded so far for Sirius A \citep{petit11,kochukhov14} display a similar shape using three different instrumental setups (ESPaDOnS, NARVAL, HARPSpol) and three different models of CCD detector and two different reduction pipelines, giving strong confidence in a stellar origin of the polarized signature. We finally emphasize that a number of stars belonging to several classes were previously observed at a comparable S/N, which resulted in no Stokes V detection in two normal B stars \citep{wade14, neiner14}, in a definite Stokes V detection (with a standard Zeeman shape) for the $\lambda$~Boo star Vega \citep{lignieres09}, and in a definite Zeeman detection (again with a standard shape) for the cool giant Pollux \citep{auriere09}.

Another potential source of instrumental artifacts, especially for very weak Stokes V signatures, is possible crosstalk from linear to circular polarization. This effect is documented for NARVAL and ESPaDOnS (e.g., \citealt{silvester12}). Stokes Q and U spectra were obtained for Sirius~A by \cite{petit11}, featuring no polarimetric signal at a level that could significantly contribute to the Stokes V signal. The same profile shape obtained for Sirius A using three instruments affected by different crosstalk levels is, in itself, an independent evidence that linear polarization did not contaminate the Stokes V signature.  

Considering this context as a whole, we conclude that a convincing body of evidence now exists to safely conclude that the Stokes V signal observed for \betuma\ and \tetleo\   most likely has a stellar origin.

\subsection{Establishing the Zeeman origin of Stokes V signatures}

\subsubsection{Method outline}

The shapes of the signatures in the Stokes V profiles (mainly constituted of a positive lobe) are not expected in the standard theory of the Zeeman effect, which predicts that lobes of positive and negative signs should be observed, resulting in a zero-integral Stokes V profile. This surprising observation, and the extremely low amplitude of the recorded signatures, raise natural concerns about possible artifacts that may contribute to the observed polarized signal. Considered all together, the standard series of tests detailed in Sect. \ref{sect:complete} provides us with strong evidence that the recorded signatures are stellar in origin. Other convincing evidence includes the possibility that crosstalk from linear to circular polarization is not involved \citep{petit11} and that no similar signatures were observed in other hot or tepid stars studied at a similar level of accuracy \citep{lignieres09,wade14,neiner14}, in spite of a strictly identical instrumental setup. HARPSpol observations reported by \cite{kochukhov14} also confirm that the peculiar signature reported for Sirius~A is still obtained when using a completely different instrument and different reduction software.   

Even if instrumental effects can be safely excluded, the physical origin of the signal still requires further investigation. We propose here a series of tests to ascertain the Zeeman origin of the recorded signal. The basic idea is that the amplitude of Zeeman signatures is expected to depend on various line parameters (Land\'e factor, wavelength, line depth), so that a careful selection of spectral lines for the LSD procedure should confirm or refute this dependence in our data. We therefore run again the LSD process using a number of new line lists, extracted from our original list but featuring a selection of lines where one line parameter has been restricted to a given range. In the weak field approximation, Stokes V signals are related to line parameters according to the following equation:

\begin{equation}
V \propto g.\lambda_0^2 .B_\parallel .\partial I / \partial \lambda
\end{equation} 

\begin{figure}[H]
\includegraphics[scale=0.34]{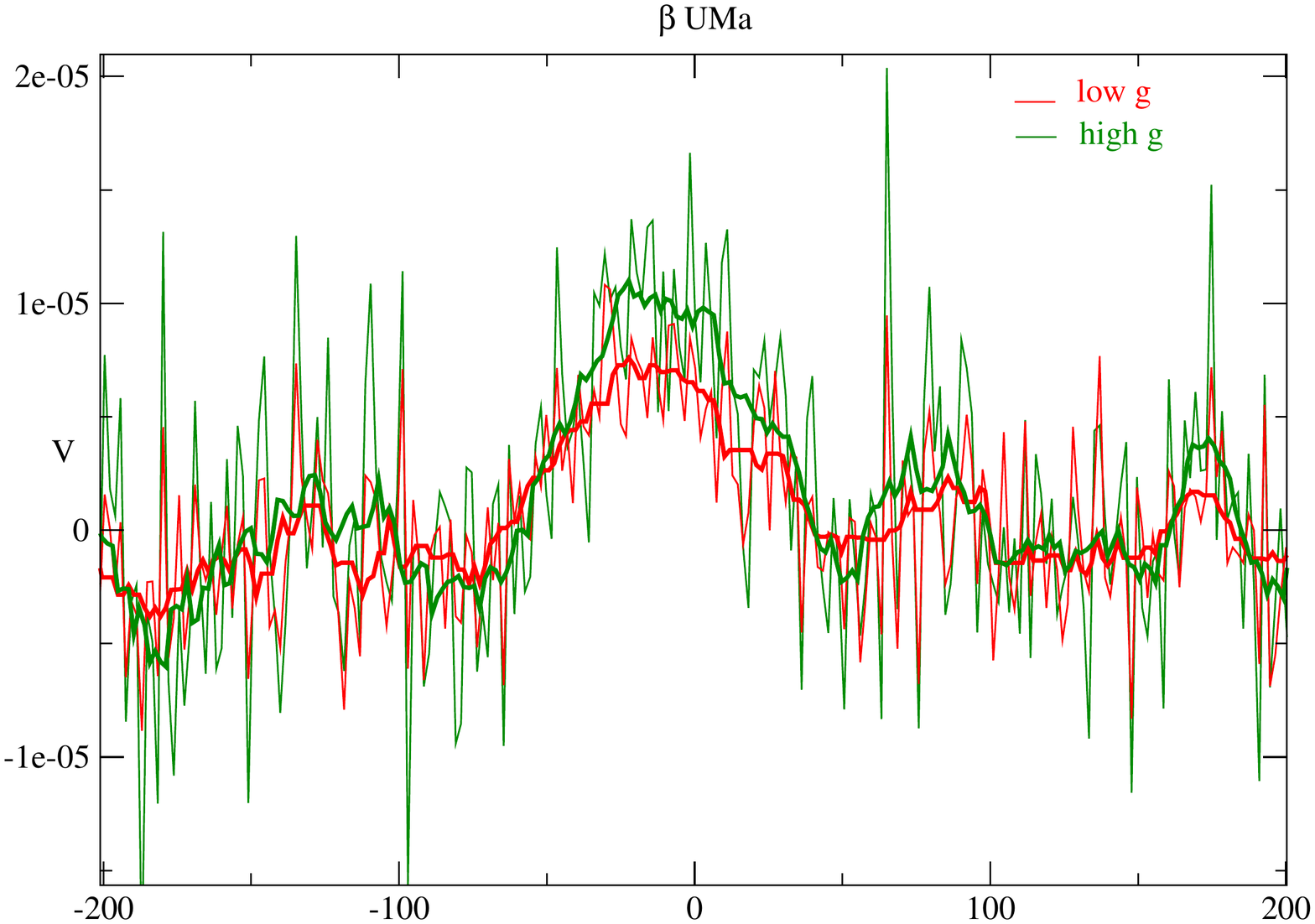}
\includegraphics[scale=0.34]{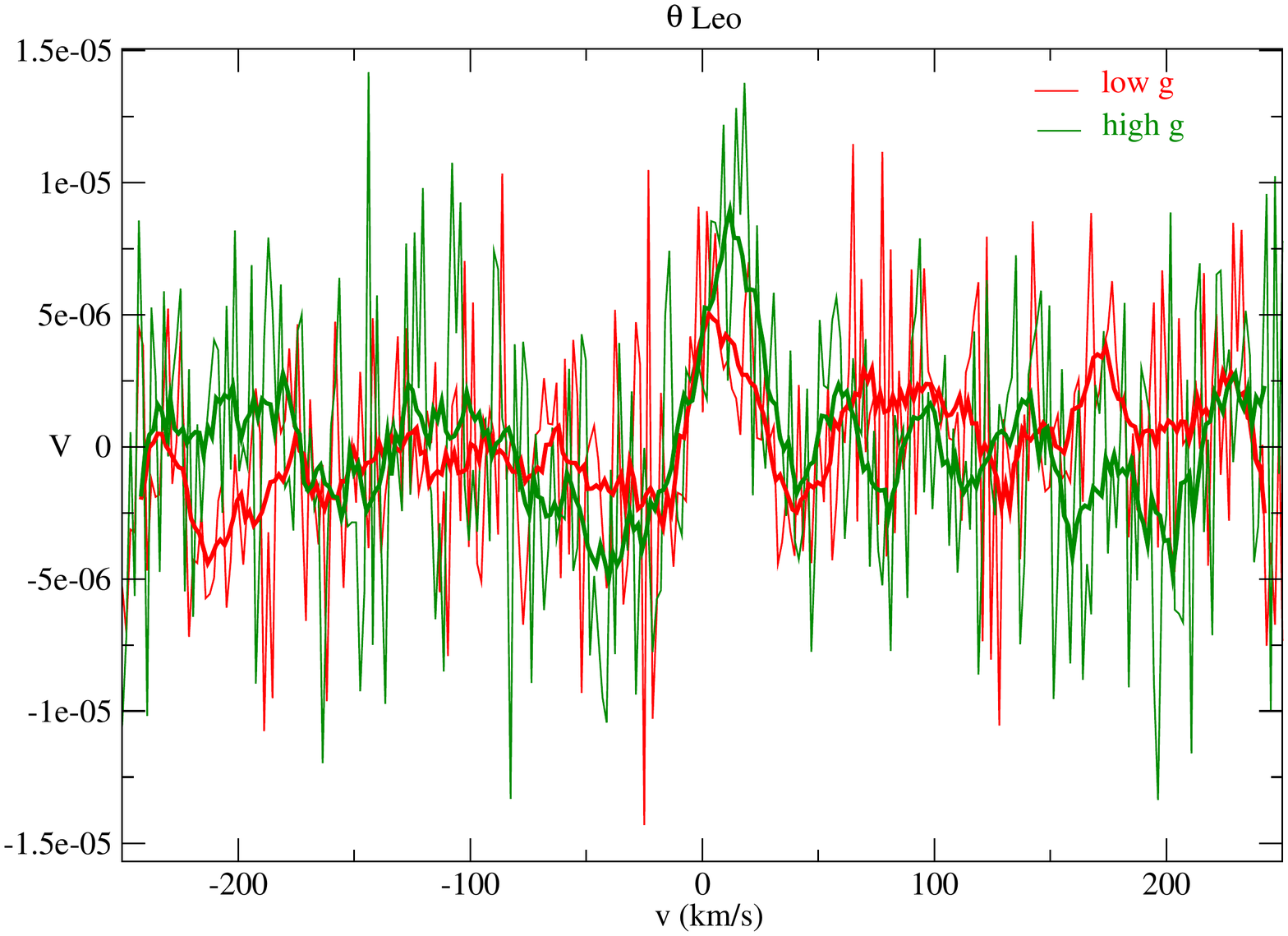}
\includegraphics[scale=0.34]{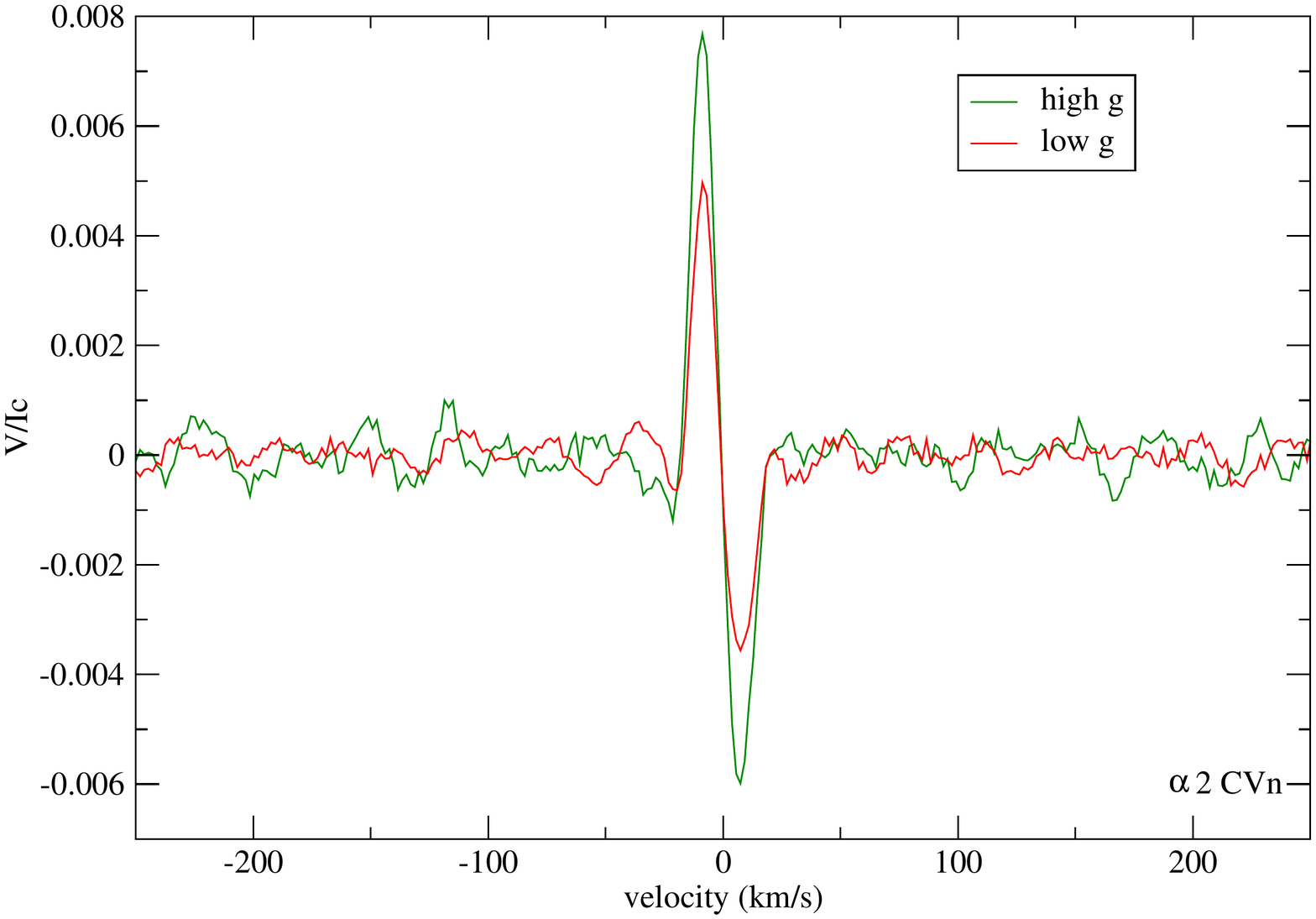}
\caption{Top: Comparison of the Stokes V profiles obtained by selecting photospheric lines of low (red thin line) and high (green thin line) magnetic sensitivity for $\beta$ UMa. The thick red and green lines represent a moving average over three spectral bins of the thin lines. Center: same figure for $\theta$ Leo. Bottom: same for $\alpha^2$ CVn. All profiles are normalized to the continuum level.}
\label{fac_lande}
\end{figure}

\noindent where $\lambda_0$ represents the wavelength of the line profile, $B_\parallel$ the line-of-sight projection of the magnetic field vector, and $g$ the effective Land\'e factor. At a given value of $B_\parallel$, the amplitude of Stokes V is therefore expected to follow simple variations with $\lambda_0$, $g$, and with the line depth.

As a reference, we use here the standard Ap star $\alpha^2$ CVn and a NARVAL observation downloaded from PolarBase \citep{petit14} and already used by \cite{silvester14}. The star  $\alpha^2$ CVn is bright and variable A0p with \vsini=$18\pm 0.5$ \kms, an effective temperature of $11600 \pm 500$ K, and a logarithmic surface gravity equal to $3.9 \pm 0.1$ \citep{silvester14}. Its spectral properties are therefore reasonably similar to $\beta$ UMa and $\theta$ Leo, except its slightly higher surface temperature. The interesting characteristic of $\alpha^2$ CVn is its strong and organized surface magnetic field (locally up to 2 kG), resulting in very large circularly polarized signatures. We applied our series of tests to this reference star to better highlight the expected results in the presence of a strong magnetic field, with negligible noise in the polarized profile. 

The average line parameters for all submasks used to compute the new LSD profiles are listed in Table \ref{lsd_parameter}. They vary slightly from one star to the next mostly because of the different VALD models employed. The largest star-to-star differences are observed when we define the line sublists according to a wavelength threshold. We also list in Table \ref{lsd_parameter} the normalization parameters used for the LSD procedure, forcing a normalized wavelength of 500~nm everywhere, except when we set a wavelength threshold, in which case we force a normalized Land\'e factor equal to 1.2. Finally, we correct for any difference in the depth of Stokes I profiles, except when the submasks are defined with a line depth threshold.

\subsubsection{Outcome for \betuma\ and \tetleo}

As a first test, we ran LSD for two submasks containing lines with an average Land\'e factor $g$ lower (resp. greater) than the mean Land\'e factor of the original line list (see Table~\ref{lsd_parameter}). Hereafter, we consider  the normalizing Land\'e factors used as part of the LSD procedure, since it is the relevant quantity for direct comparison of different LSD profiles. (The normalizing $g$ values follow the same trend as the average Land\'e factors of the submasks.) The resulting Stokes V profiles are plotted in Fig. \ref{fac_lande} for the two Am stars and the control Ap star. The Stokes V profiles are corrected for a $\sim$10\% difference in equivalent width observed in their associated Stokes I profile. Because of a higher noise level than obtained with the complete line mask, the high-$g$ and low-$g$ profiles of $\beta$ UMa and $\theta$ Leo do not display any statistically conclusive differences. The overplotted running average helps to improve the situation, showing that the high-$g$ signals possess higher amplitudes than their low-$g$ counterparts. We note that their amplitude ratio is roughly consistent with the g ratio, although this point is difficult to establish with high accuracy (even with the running average) because of the level of noise.

\begin{figure}[H]
\includegraphics[scale=0.34]{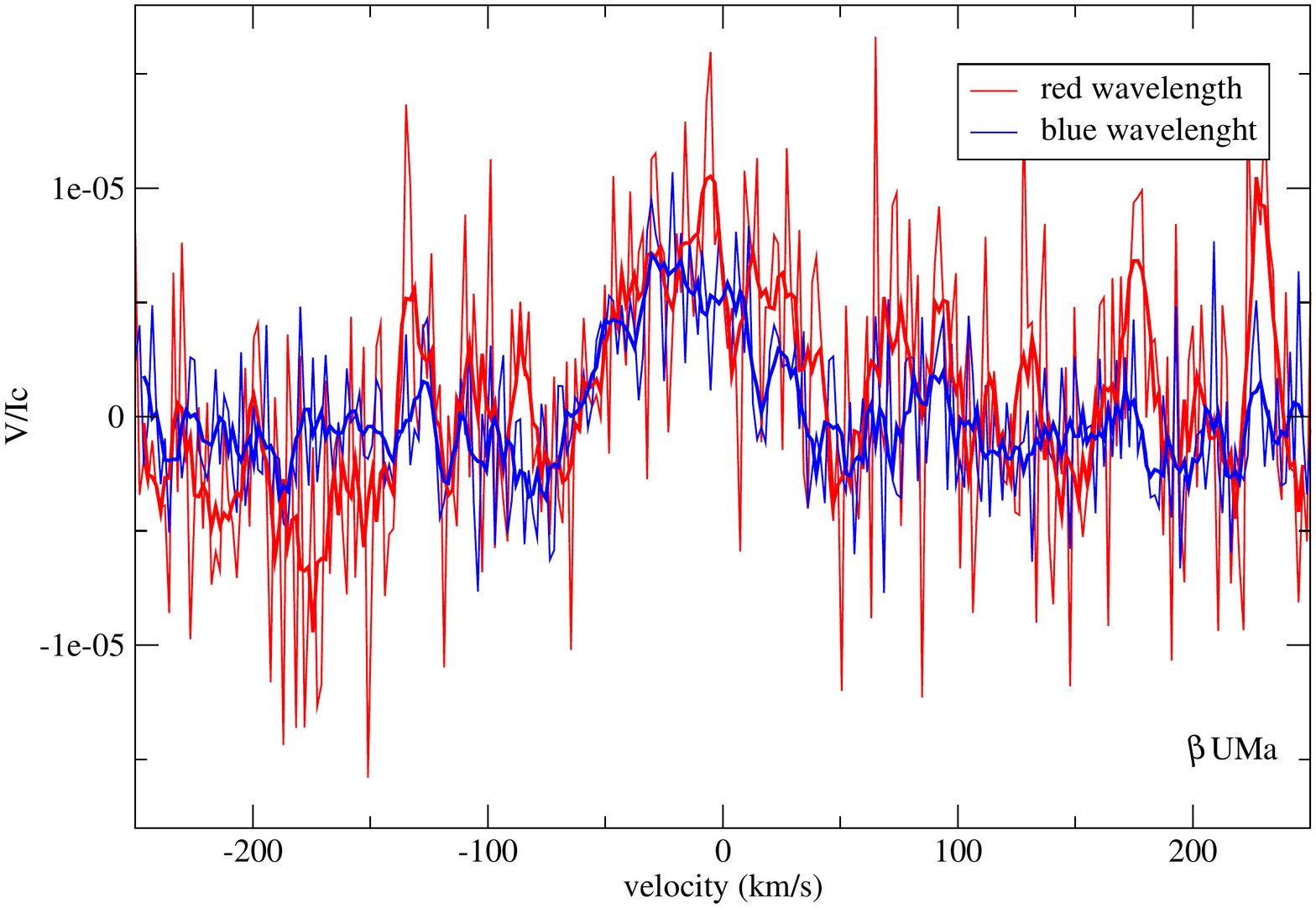}
\includegraphics[scale=0.34]{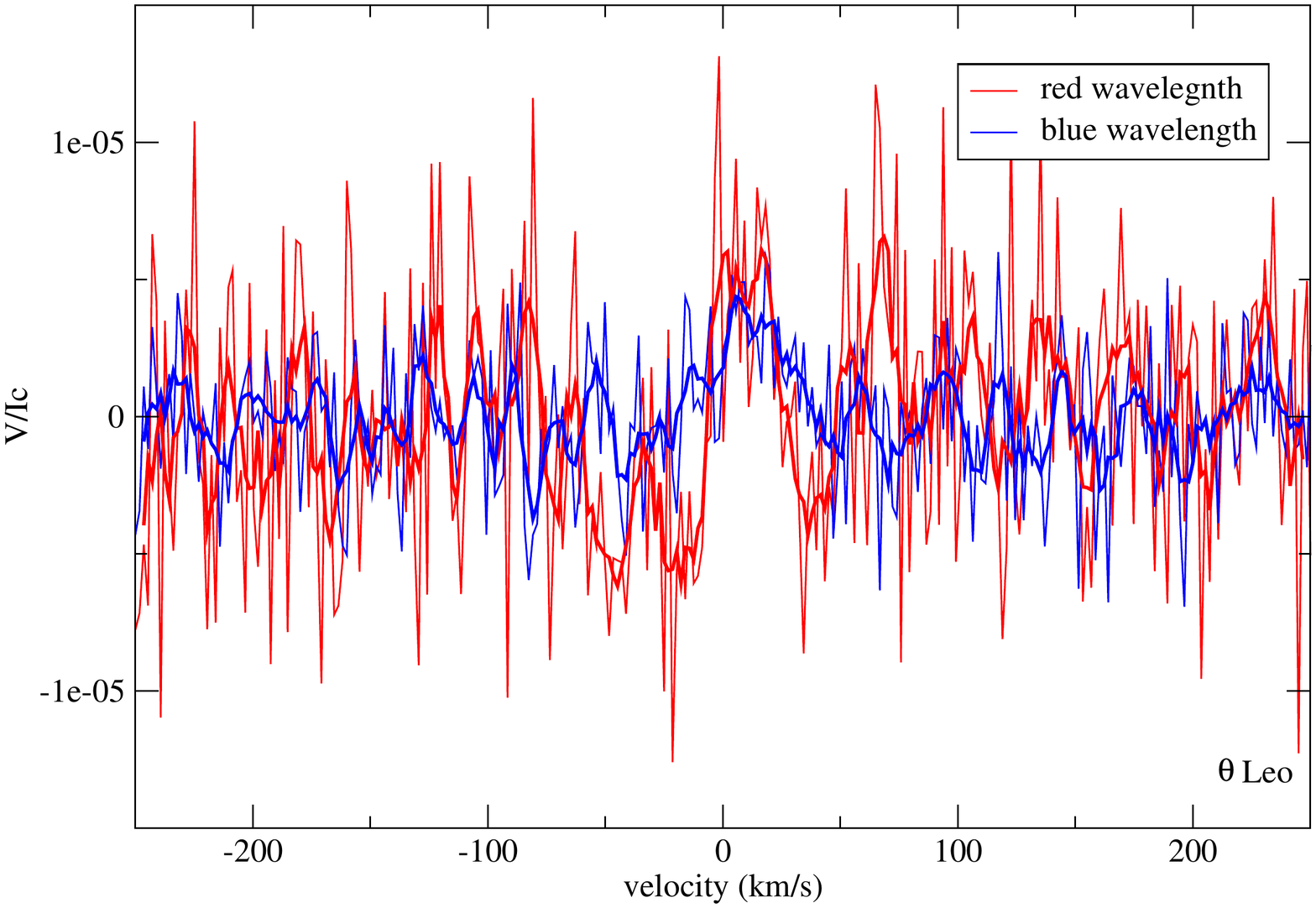}
\includegraphics[scale=0.34]{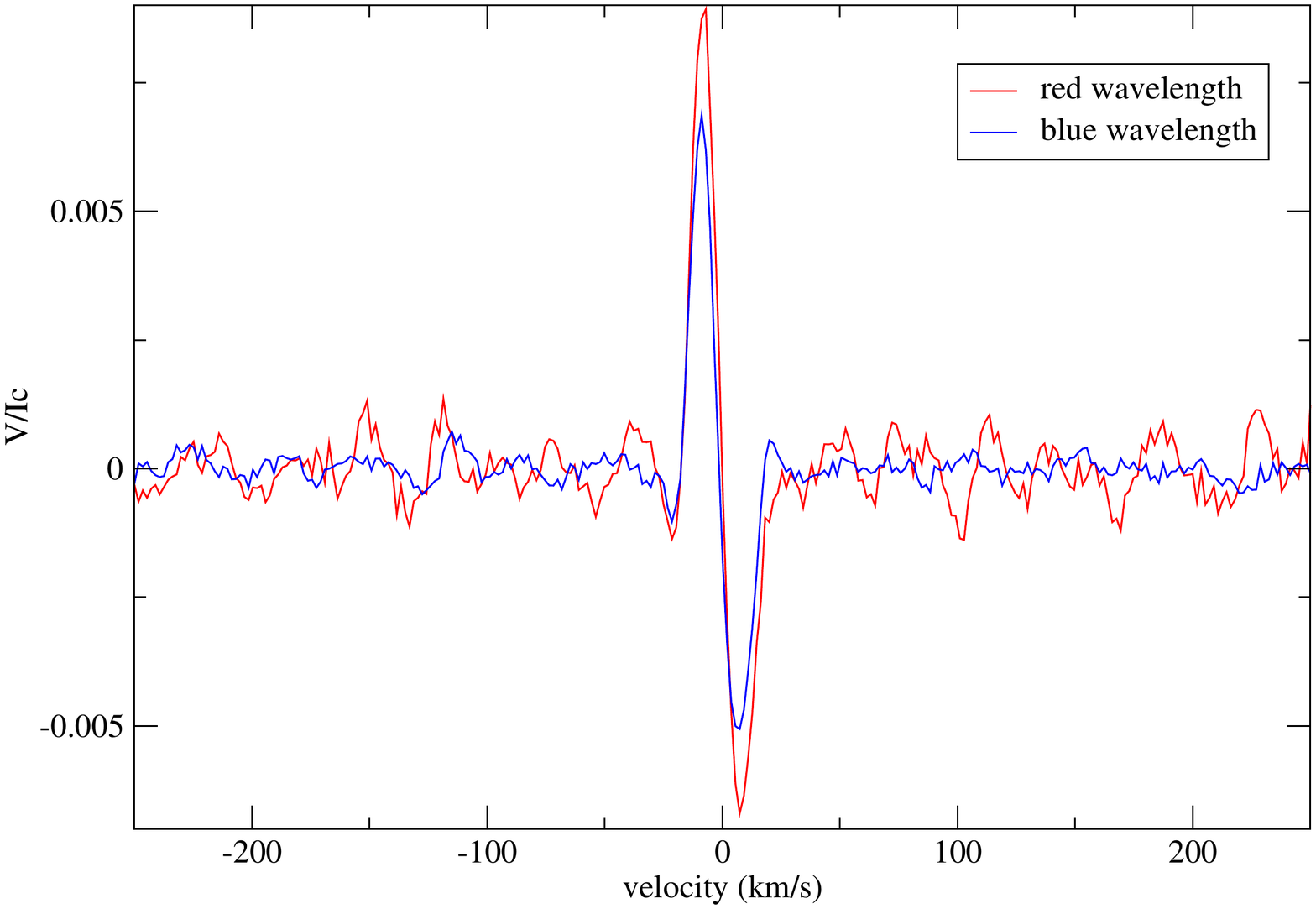}
\caption{Same as Fig. \ref{fac_lande} for photospheric lines of low (thin red line) and high (thin green line) wavelength.}
\label{wave}
\end{figure}

As second test, two sublists were defined from our original list by containing lines with a wavelength lower (resp. greater) than the mean wavelength of the original list (Table~\ref{lsd_parameter}). For a given star, the Stokes V profiles were corrected for the $\sim$30\% difference in equivalent width observed in their associated Stokes I profiles. The outcome of this test is illustrated in Fig.~\ref{wave}, which illustrates a marginally larger amplitude of the Stokes V signal when the wavelength increases. As for the previous test, we computed a moving average of the signal to confirm the trend that is otherwise completely hidden in the noise and to check that the trend observed in both Am stars is consistent with the outcome obtained for $\alpha$ 2 CVn.
 
\begin{figure}[H]
\includegraphics[scale=0.34]{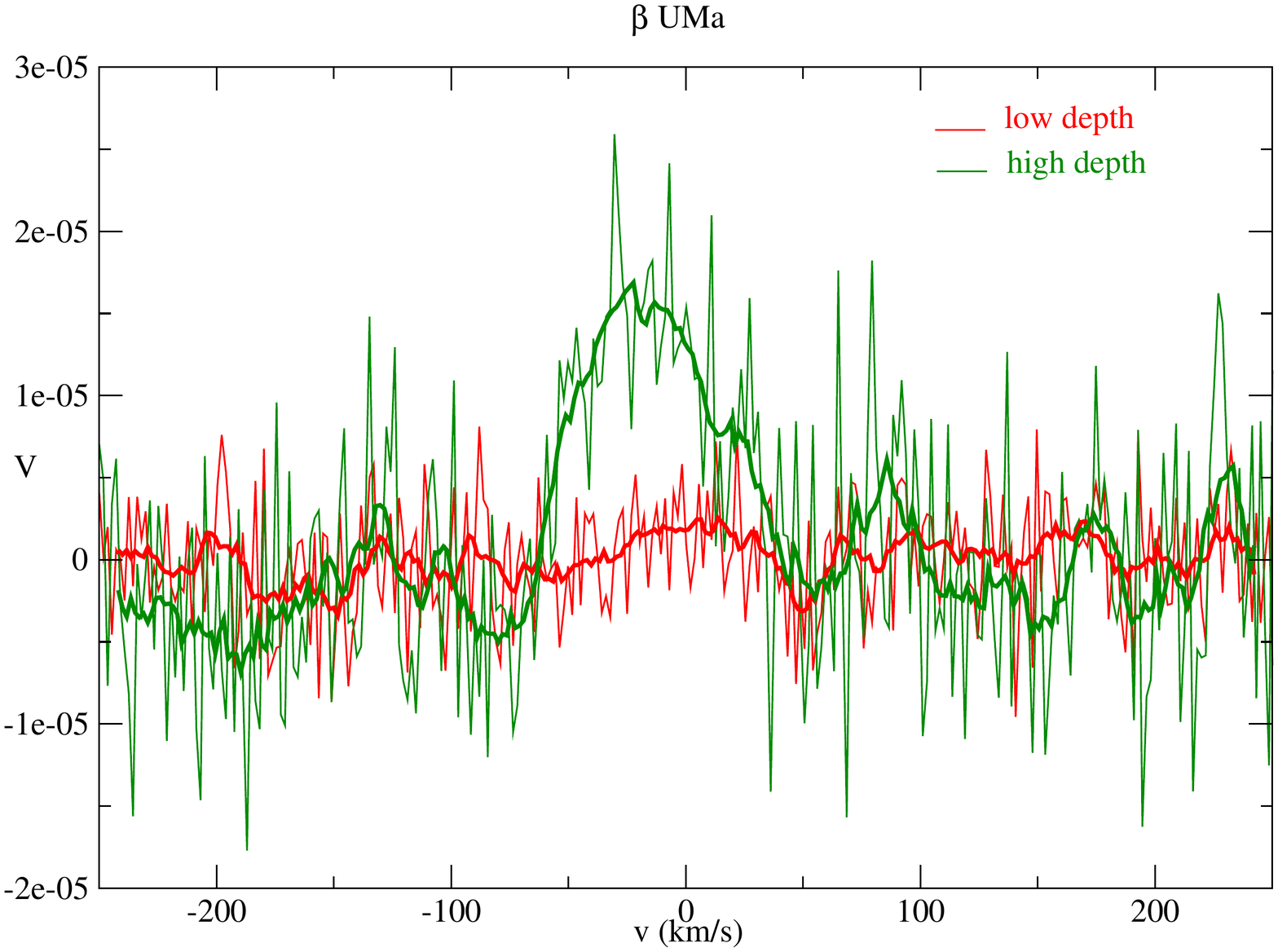}
\includegraphics[scale=0.34]{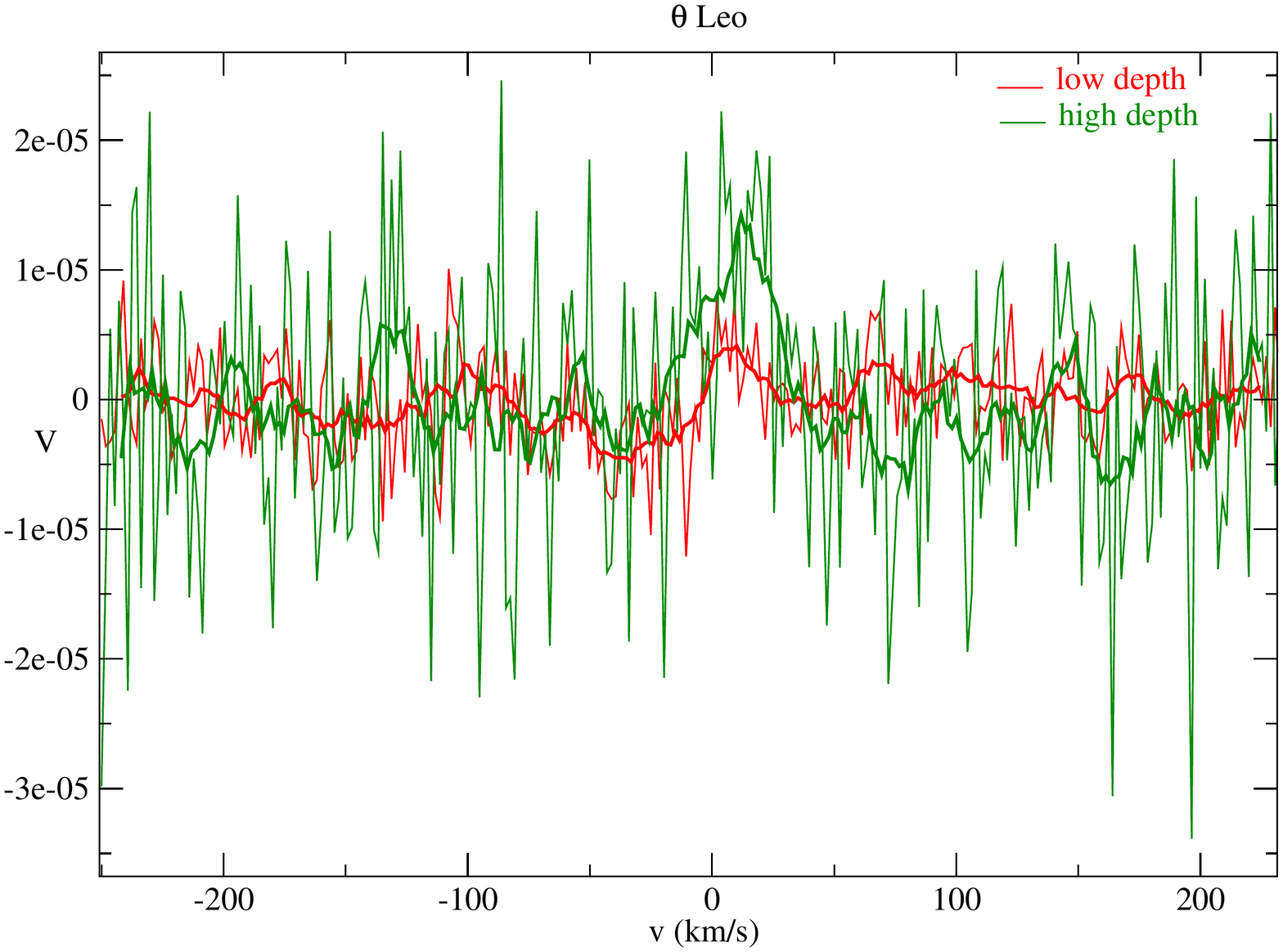}
\includegraphics[scale=0.34]{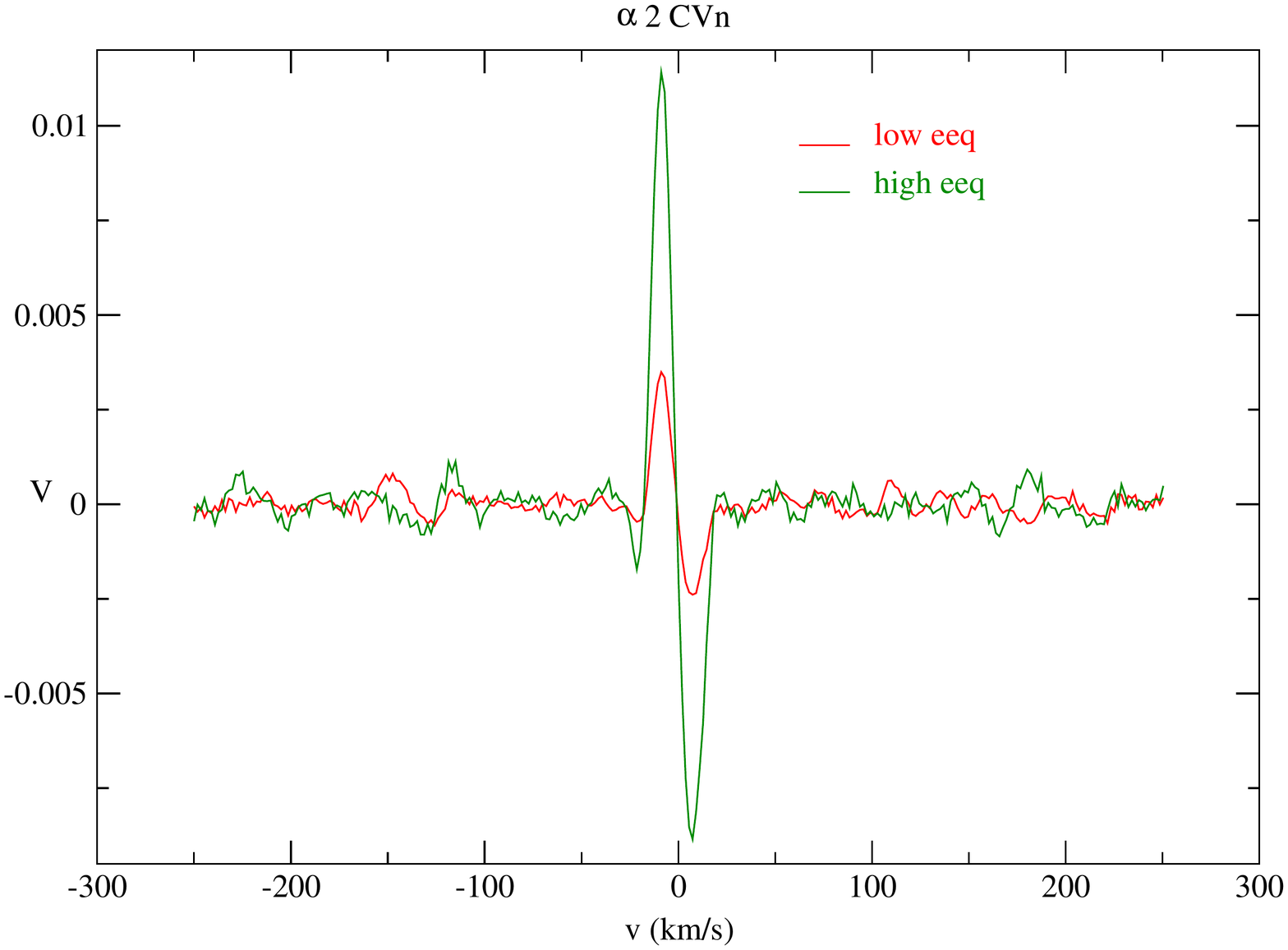}
\caption{Same as Fig. \ref{fac_lande} for photospheric lines of low (thin red line) and high (thin green line) central depth.}
\label{eeq}
\end{figure}

As a last test, we define two sublists using spectral lines with an average depth lower (resp. greater) than the mean depth of the original list (Table~\ref{lsd_parameter}). The Stokes V LSD profiles obtained from the sublists are displayed in Fig.~\ref{eeq} and, for our two Am stars and our reference star, clearly show a lower amplitude whenever the average line depth is smaller. This outcome is expected in the case of a signature of magnetic origin, but also for most instrumental artifacts. 

From the series of tests presented here, only the one with low versus high Land\'e factors was performed by \cite{petit11} for Sirius~A. For consistency, we used their observing material to reproduce with Sirius~A the three tests applied to $\beta$ UMa and $\theta$ Leo. The result, not shown here, is fully consistent with the conclusions reached in the present study. Considered together, this series of tests strongly suggests that the circularly polarized signatures obtained for the three bright Am stars observed so far have a Zeeman origin.

Based on this conclusion, it is tempting to estimate the surface field strength from our set of measurements, using the classical center of gravity (or first moment) method \citep{rees79}. We must stress, however, that this widely used technique is based on the standard assumption that the Stokes V signature is antisymmetric about the line center, which is very far from the actual shape of our Stokes V signatures. A purely symmetric signature (closer to what is obtained for $\beta$ UMa and $\theta$ Leo) will be interpreted as a zero longitudinal field strength, regardless of the amplitude of the Stokes V signal, similarly to dipolar fields observed at the rotational phase of a crossover configuration (e.g., \citealt{auriere07}). The situation here is obviously different, because the large time span of data collection is very unlikely to be restricted to a crossover phase. Nevertheless, such a measurement (and in particular its error bar) provides us with a quality measure of the sensitivity of the magnetic diagnosis that can be compared to similar studies. The first moment estimate of the magnetic field provides us with a field strength of $-1\pm0.8$~G for $\beta$ UMa and $-0.4\pm0.3$~G for $\theta$ Leo that is unsurprisingly consistent with zero (as previously reported with Sirius~A). As an attempt to propose a more relevant proxy of the field strength, we calculate the equivalent width (EW) of the Stokes V signature and normalize this EW by the one of the Stokes I profile. By doing so, we obtain a normalized EW equal to $1.96 \times 10^{-4}$ for $\beta$ UMa, and $5.44 \times 10^{-5}$ for $\theta$ Leo. For Sirius~A, the normalized EW is equal to $6.68\times 10^{-5}$.   

\section{Discussion}

\subsection{Peculiar Stokes V signatures in Am stars}

The observations presented here provide new clues to the weak polarized signatures produced in the photospheres of intermediate-mass stars. We report the detection of weak Stokes V signatures in two of the brightest Am stars, which complements the previous detection of a similar polarized signal for Sirius~A \citep{petit11}\footnote{an observation confirmed by independent HARPSpol observations carried out by \cite{kochukhov14}.}. Considered together, the three polarimetric detections constitute a 100 percent detection rate so far in our sample of bright Am stars, suggesting widespread similar signatures in this stellar class. 

All signals observed to now possess roughly the same shape, with one positive lobe roughly symmetric about the line center and occupying most of the width of the line profile. Negative lobes surrounding the positive one, if they exist, do not exceed the noise level in the data sets available to us. These profile shapes displaying net circular polarization are atypical of Zeeman signatures observed in other classes of magnetic stars, where the integral of the Stokes V profile is generally close to null. This peculiar shape naturally prompts questions about the origin of these polarized spectral features. The tests conducted in our study show that these unexpected signatures depend on spectral line parameters (wavelength, Land\'e factor, line depth), as expected from a Zeeman signal. 

Stokes V profiles that are nearly symmetric about the line center are common. These patterns are temporarily observed when two magnetic poles of an inclined dipole are seen on the visible hemisphere of a star, close to the limb, and therefore with different radial velocities (at the so-called crossover rotational phases). This simple interpretation is, however, very unlikely here since our co-added LSD profiles mix data collected over timespans much longer than the typical rotation periods of Am stars, merging a large number of random rotation phases.  A dominant toroidal magnetic field component is also able to generate symmetric Stokes V profiles \citep{donati05}, although this specific type of magnetic geometry should not produce any net circular polarization, as observed here. In any case, a purely geometric explanation is not able to account for the absence of negative lobes in the Stokes V profiles.    

A number of cool active stars were reported to display weak net circular polarization after integration over LSD line profiles \citep{petit05, auriere08, morgenthaler12, auriere11, tsvetkova13, lebre14}. However, no similar findings have been reported so far in strongly magnetic massive stars or intermediate-mass stars, and the very subtle effect reported for cool stars is nowhere near the extreme stituation reported here. For cool stars, the proposed interpretation was adapted from solar physics, where abnormal Stokes V are routinely described (e.g., \citealt{solanki93}) and attributed to simultaneous vertical gradients in velocities and magnetic field strengths (\citealt{lopez02} and references therein). Single-lobed signatures resembling those recorded for Am stars can be locally observed in solar magnetic elements \citep{viticchie11,sainz12}, but they are more difficult to justify in the case of disk-integrated measurements (as obtained for unresolved stars) because of the organized flows and magnetic fields invoked to justify their shape. Relatively strong magnetic fields are also involved in asymmetric solar Stokes V profiles, although the very weak disk-integrated signatures reported here do not tell much about local magnetic strengths, which could potentially be rather large in the case of a very tangled field geometry.

The absence of any similar phenomenon in Ap stars (in spite of masses roughly identical to those of Am stars) may simply be related to the lack of any significant surface turbulence due to the strong magnetic fields permeating their photosphere \citep{folsom13} and, in the case of Bp stars, to a photospheric temperature too high to allow for a thin convective shell, even in the absence of their magnetic field. The situation is different for Am stars, for which high-resolution spectra have revealed stronger microturbulence than for normal A stars \citep{landstreet09}, as long as their effective temperature remains below about 10,000~K, a condition fulfilled by our two targets and by Sirius~A. The very shallow convective shell producing this turbulent velocity field may host supersonic convection flows \citep{kupka09}. This could provide the source of sharp velocity and magnetic gradients needed to produce strongly asymmetric profiles. Shocks traveling in this superficial turbulent zone may also contribute to amplify any existing magnetic field, as previously proposed in the context of the Mira star $\chi$~Cygni \citep{lebre14}. 

In any case, a physical model able to produce a convincing reproduction of the peculiar polarized signatures reported for Am stars still needs to be developed. Preliminary simulations of Stokes V profiles with
velocity and field gradients show that signatures such as those observed
in beta UMa, theta Leo, and Sirius A can be reproduced (C. Folsom, priv.
comm.). Without such a tool at our disposal, any quantitative description of the associated surface magnetic fields is out of reach, since techniques commonly used to estimate stellar magnetic field strengths (like the center-of-gravity method) are not suited to model Stokes V profiles following such unexpected shape. In practice, magnetic strengths derived for \betuma\ and \tetleo\ by applying the usual methods can only provide us with a lower limit of a few tenths of a gauss on the surface axisymmetric field component, which is consistent with the estimate available for on Sirius~A.

\subsection{Origin of the magnetism of Am stars}

\begin{figure}
\includegraphics[scale=0.34]{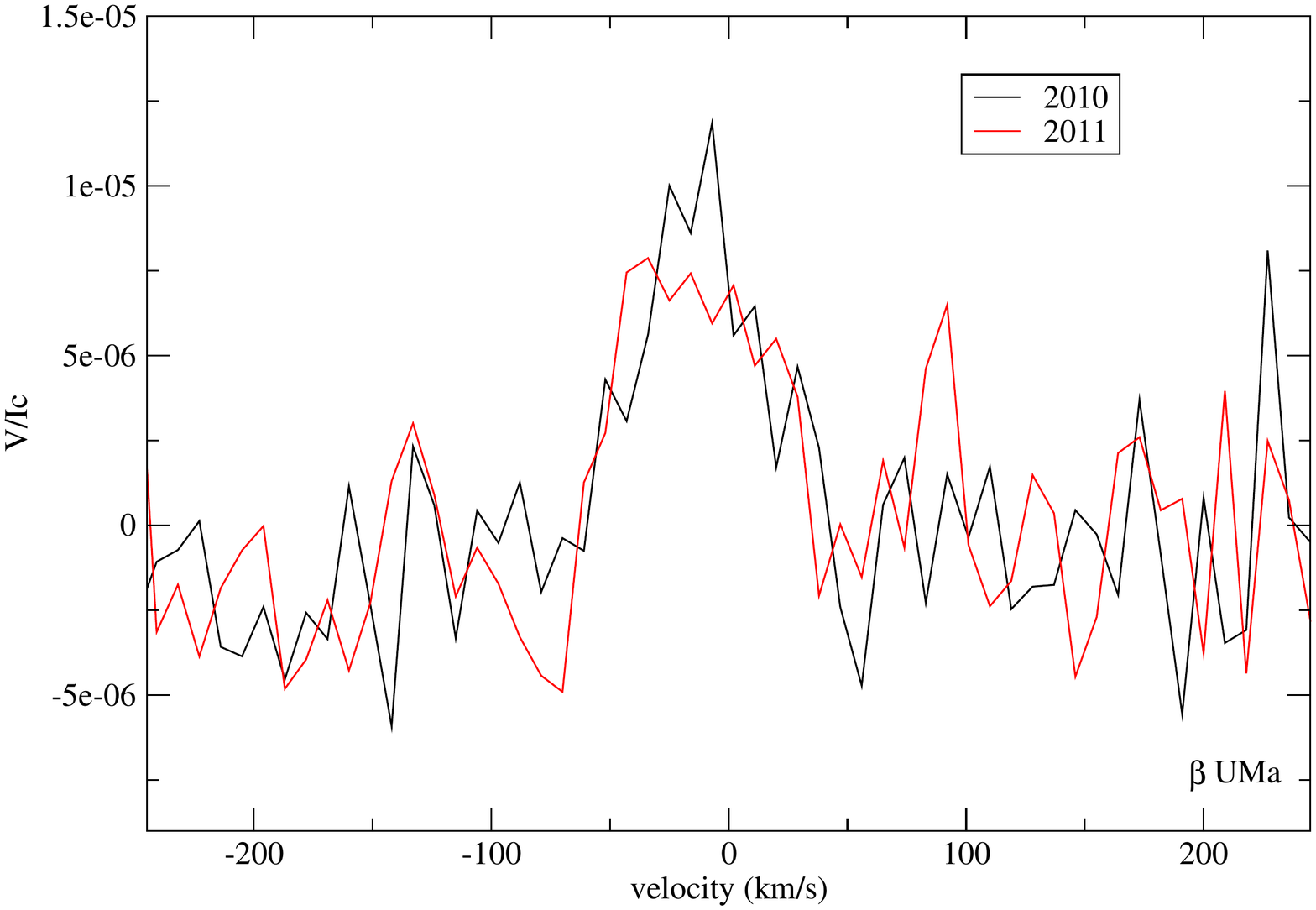}
à\includegraphics[scale=0.34]{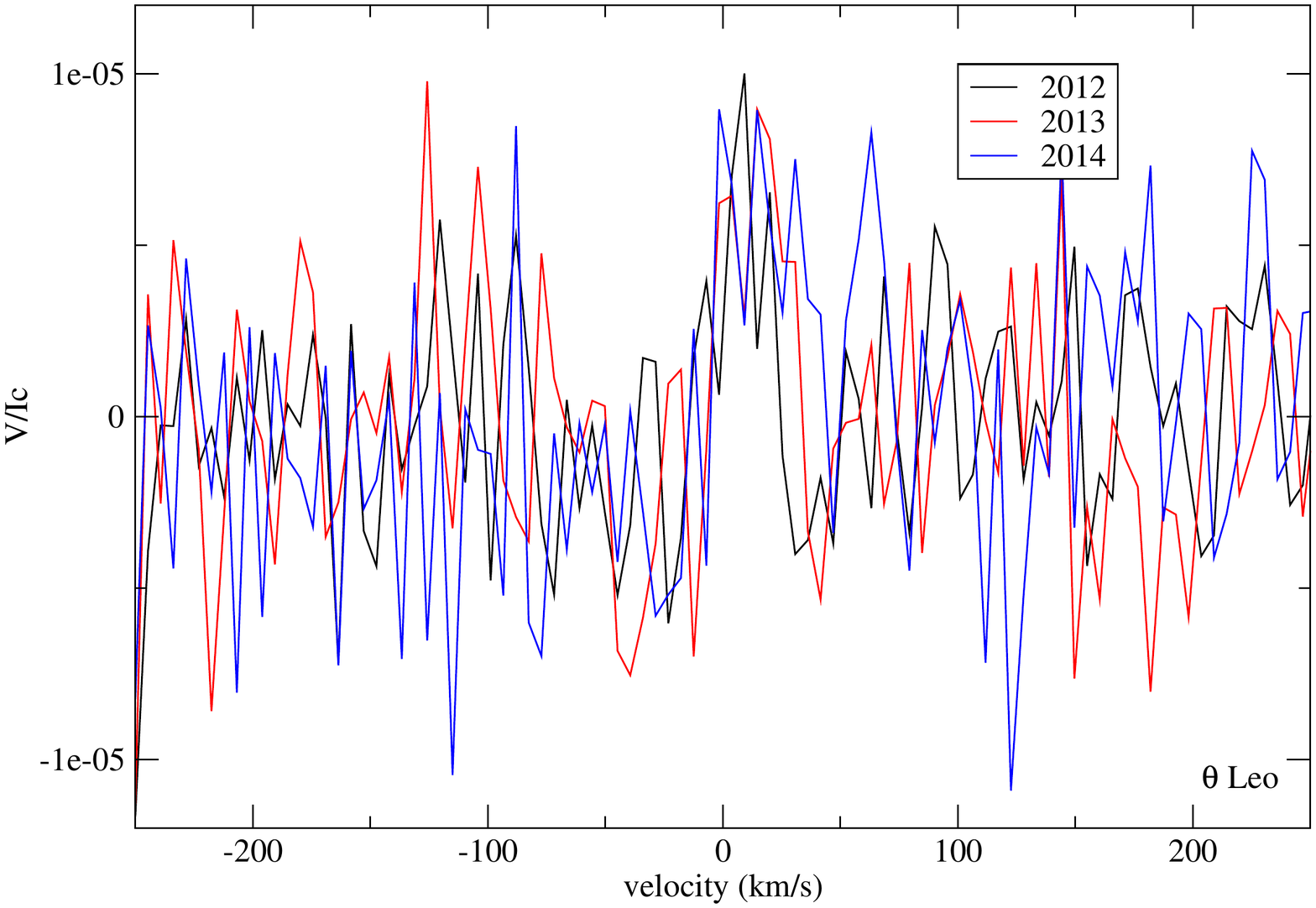}
\caption{Top: Mean Stokes V profiles for the different years of observation of $\beta$ UMa. Bottom: same figure for $\theta$ Leo.}
\label{year}
\end{figure}

With only small number of objects observed so far and polarimetric signatures close to the detection limit, our observations only offer a few hints to the physical ingredients involved in the generation of the weak surface magnetic fields observed in Am stars.

An important clue to distinguishing between a dynamo-generated field and most other scenarios is the long-term evolution of the observed magnetic field, because a dynamo-generated field is likely to experience some temporal variability on a secular timescale. By splitting our data sets into subsets limited to a given year of observation, we are able to get a first glimpse at the stability of the polarimetric signal (Fig.~\ref{year}). We find that signatures recovered one year apart are consistent with each other, showing that any variability over this timespan remains below the noise level. This outcome is consistent with similar attempts for Sirius~A and Vega.

Surface brightness inhomogeneities are usually associated with the structured magnetic field produced by a global dynamo. The lack of any rotation period estimate available in the literature for \betuma\ and \tetleo\ suggests that any such brightness patches must take place on relatively small spatial scales or be limited to a very low contrast. We note that the recent discovery of rotational modulation in Am stars of the Kepler field by \cite{balona15} was limited to targets that are significantly cooler than the objects of our study. A possibility is that the deeper convective envelope of stars in the Kepler sample may be more favorable to the onset of a large-scale dynamo. A very sensitive method, such as the one employed by \cite{bohm15} to detect very faint starspots on Vega, may be the key to unveiling surface features on weakly magnetic Am stars like those studied here. We also note the lack of documented flaring events for these bright and well-studied stars, again in contrast to claims for cooler intermediate-mass stars observed with Kepler \citep{balona13,balona15}.

\subsection{Toward a systematic exploration of weak magnetic fields in Am stars}

The ultra-deep polarimetric campaign carried out for three bright Am stars is far from exhausting the exploration of this stellar class. The most noticeable difference between these stars is that \betuma\ and \tetleo\ seem to be located near the end of the main sequence, while Sirius~A is a more standard main sequence object. One conclusion of our study is that these differences in the evolutionary status do not affect the recorded polarimetric signatures in any obvious way.

While Sirius~A and \tetleo\ share a low projected rotational velocity, \betuma\ displays a higher \vsini\ value, although it is not possible to distinguish between the contribution of rotational and inclination effects in this parameter. The larger normalized EW of the polarimetric signal reported for \betuma\ may be a first hint of a rotational dependence of the weak magnetism of Am stars, although a much larger sample is required to seriously test this hypothesis.
 
Finally, all three objects observed so far were confined to a quite narrow band in effective temperature. The active behavior of cooler Am stars \citep{balona13,balona15} is a strong motivation to expand the available sample to Am stars of late-A spectral types. Since the peculiar polarized signatures observed up to now are proposed to be indirect tracers of surface convective motions,
gathering observations in cooler stars is an obvious way to test this hypothesis by considering the effects of varying the surface turbulent flows on the polarized signature.

\begin{acknowledgements}
We acknowledge support from the ANR (Agence Nationale de
la Recherche) project Imagine. This research has made use of the SIMBAD database
operated at the CDS, Strasbourg (France) and of NASA's Astrophysics Data System
(ADS). CPF is supported by the ANR grant "Toupies: Towards understanding the spin evolution of stars". GAW acknowledges Discovery Grant support from the Natural Science and Engineering Research Council (NSERC) of Canada. We are grateful to the referee, Iosif Romanyuk, for constructive comments.
\end{acknowledgements}

\bibliographystyle{aa} 
\bibliography{biblio_v2}

\Online

\begin{appendix}

\section{Journal of observations}

\begin{table}[h]
\label{obs}
\caption{Journal of observations. The columns contain the date for
each Stokes V sequence, the heliocentric Julian date corresponding to
the middle of the observation time, the object name, the number of
sequences, the exposure time per individual subexposure and the
averaged S/N in the individual LSD Stokes V
pseudo-line profiles ($\pm$rms).}
\centering
\tabcolsep=3pt
\begin{tabular}{c c c c c}
\hline
Date & Mid-HJD & Star & $T_{exp}$ (s) & S/N\\
\hline
\hline
17 Mar 2010 & 2455273.520 & $\beta$ UMa & 16$\times$4$\times$107 &
52707$\pm$6090\\
06 Apr 2010 & 2455293.412 & $\beta$ UMa & 17$\times$4$\times$107 &
49436$\pm$22333\\
10 Apr 2010 & 2455297.444 & $\beta$ UMa & 19$\times$4$\times$107 &
76493$\pm$1960\\
11 Apr 2010 & 2455298.397 & $\beta$ UMa & 19$\times$4$\times$107 &
32500$\pm$7752\\
25 Mar 2011 & 2455646.426 & $\beta$ UMa & 25$\times$4$\times$107 &
56378$\pm$24444\\
31 Mar 2011 & 2455652.504 & $\beta$ UMa & 25$\times$4$\times$107 &
45963$\pm$4907\\
02 Apr 2011 & 2455654.379 & $\beta$ UMa & 03$\times$4$\times$107 &
53964$\pm$8998\\
04 Apr 2011 & 2455656.462 & $\beta$ UMa & 24$\times$4$\times$107 &
69029$\pm$6099\\
22 Jan 2012 & 2455949.644 & $\theta$ Leo & 05$\times$4$\times$180 &
44503$\pm$1018\\
23 Jan 2012 & 2455950.628 & $\theta$ Leo & 05$\times$4$\times$180 &
39774$\pm$5090\\
24 Jan 2012 & 2455951.624 & $\theta$ Leo & 05$\times$4$\times$180 &
41547$\pm$3889\\
25 Jan 2012 & 2455952.640 & $\theta$ Leo & 05$\times$4$\times$180 &
41737$\pm$3134\\
14 Mar 2012 & 2456001.579 & $\theta$ Leo & 05$\times$4$\times$180 &
43929$\pm$1810\\
15 Mar 2012 & 2456002.524 & $\theta$ Leo & 10$\times$4$\times$180 &
47360$\pm$698\\
24 Mar 2012 & 2456011.526 & $\theta$ Leo & 05$\times$4$\times$180 &
44880$\pm$1487\\
25 Mar 2012 & 2456012.502 & $\theta$ Leo & 05$\times$4$\times$180 &
47392$\pm$506\\
27 Mar 2012 & 2456013.400 & $\theta$ Leo & 10$\times$4$\times$180 &
40883$\pm$1229\\
21 Mar 2013 & 2456373.488 & $\theta$ Leo & 09$\times$4$\times$180 &
25542$\pm$4619\\
23 Mar 2013 & 2456375.465 & $\theta$ Leo & 09$\times$4$\times$180 &
29220$\pm$2557\\
16 Apr 2013 & 2456399.444 & $\theta$ Leo & 09$\times$4$\times$180 &
23751$\pm$3600\\
17 Apr 2013 & 2456400.492 & $\theta$ Leo & 09$\times$4$\times$180 &
45010$\pm$1529\\
22 Apr 2013 & 2456405.512 & $\theta$ Leo & 09$\times$4$\times$180 &
42777$\pm$1707\\
23 Apr 2013 & 2456406.454 & $\theta$ Leo & 09$\times$4$\times$180 &
42064$\pm$2815\\
24 Apr 2013 & 2456407.502 & $\theta$ Leo & 09$\times$4$\times$180 &
39578$\pm$2497\\
14 Apr 2014 & 2456762.445 & $\theta$ Leo & 05$\times$4$\times$180 &
25433$\pm$8566\\
07 May 2014 & 2456785.408 & $\theta$ Leo & 05$\times$4$\times$180 &
42839$\pm$3748\\
08 May 2014 & 2456786.411 & $\theta$ Leo & 05$\times$4$\times$180 &
39435$\pm$2842\\
09 May 2014 & 2456787.416 & $\theta$ Leo & 05$\times$4$\times$180 &
44236$\pm$617\\
14 May 2014 & 2456792.471 & $\theta$ Leo & 05$\times$4$\times$180 &
42041$\pm$543\\
15 May 2014 & 2456793.413 & $\theta$ Leo & 05$\times$4$\times$180 &
44653$\pm$1052\\
07 Jun 2014 & 2456816.408 & $\theta$ Leo & 05$\times$4$\times$180 &
29599$\pm$1530\\
10 Jun 2014 & 2456819.415 & $\theta$ Leo & 05$\times$4$\times$180 &
29931$\pm$3928\\
\hline
\end{tabular}
\end{table}

\section{Parameters of LSD profiles}

\begin{table}[h]
\caption{Mean and normalization parameters of the original mean LSD line profiles for $\beta$ UMa, $\theta$ Leo, and $\alpha^2$ CVn.}
\label{lsd_parameter}
\centering
\begin{tabular}{c c c c}
\hline
  & $\beta$ UMa & $\theta$ Leo & $\alpha^2$ CVn\\
\hline
Original Mask\\
\hline
Mean Land\'e factor \textit{g} & 1.207 & 1.206 & 1.218\\
Mean wavelength (nm) & 475.72  & 489.05 & 493.34\\
Mean line depth & 0.322 & 0.311 & 0.297 \\
Normalized Land\'e factor & 1.216 & 1.227 & 1.241\\
Normalized wavelength (nm) & 500.00 &500.00 & 500.00\\
Normalized depth & 0.450 & 0.433 &0.410 \\
\hline
Low\textit{g}LSD\\
\hline
Mean Land\'e factor\textit{g}& 0.941 & 0.956 & 0.971\\
Mean wavelength (nm) & 472.22  & 488.71 & 494.91\\
Mean line depth & 0.311 & 0.317 & 0.308\\
Normalized Land\'e factor & 0.939 & 0.957 & 0.992\\
Normalized wavelength (nm) & 500.00 &500.00 & 500.00\\
Normalized depth & 0.464 & 0.441 &0.421 \\
\hline
High\textit{g}LSD\\
\hline
Mean Land\'e factor\textit{g}& 1.529 & 1.516 & 1.533\\
Mean wavelength (nm) & 479.95  & 489.48 & 491.27\\
Mean line depth & 0.310 & 0.305 & 0.283\\
Normalized Land\'e factor & 1.469 & 1.463 & 1.489\\
Normalized wavelength (nm) & 500.00 &500.00 & 500.00\\
Normalized depth & 0.436 & 0.4253 &0.397 \\
\hline
Low wavelength LSD\\
\hline
Mean Land\'e factor\textit{g}& 1.219 & 1.217 & 1.229\\
Mean wavelength (nm) & 419.88  & 420.29 & 400.88\\
Mean & 0.381 & 0.361 & 0.049\\
Normalized Land\'e factor & 1.2 & 1.2 & 1.2\\
Normalized wavelength (nm) & 450.03 & 457.57 & 464.26\\
Normalized depth & 0.522 & 0.495 & 0.465 \\
\hline
High wavelength LSD\\
\hline
Mean Land\'e factor\textit{g}& 1.197 & 1.195 & 1.206\\
Mean wavelength (nm) & 573.3  & 604.7 & 713.43\\
Mean line depth & 0.275 & 0.267 & 0.256\\
Normalized Land\'e factor & 1.2 & 1.2 & 1.2\\
Normalized wavelength & 593.28 & 606.06 & 623.95\\
Normalized depth & 0.375 & 0.358 &0.333 \\
\hline
Low depth LSD\\
\hline
Mean Land\'e factor\textit{g} & 1.219 & 1.209 & 1.226\\
Mean wavelength (nm) & 505.58  & 507.28 & 512.475\\
Mean line depth & 0.214 & 0.212 & 0.203\\
Normalized Land\'e factor & 1.328 & 1.358 & 1.405\\
Normalized wavelength (nm) & 500.00 &500.00 & 500.00\\
Normalized depth & 0.253 & 0.252 &0.238 \\
\hline
High depth LSD\\
\hline
Mean Land\'e factor\textit{g}& 1.170 & 1.192 & 1.193\\
Mean wavelength (nm) & 480.14  & 481.97 & 480.45\\
Mean line depth & 0.628 & 0.643 & 0.588\\
Normalized Land\'e factor & 1.205 & 1.141 & 1.208\\
Normalized wavelength (nm) & 500.00 &500.00 & 500.00\\
Normalized depth & 0.646 & 0.649 &0.599 \\
\hline
\end{tabular}
\end{table}
\end{appendix}

\end{document}